\documentclass[iop]{emulateapj}

\usepackage{amssymb,amsmath}
\usepackage{graphicx}
\usepackage[percent]{overpic}
\usepackage{overpic}
\usepackage{color}
\usepackage[colorlinks=true]{hyperref}
\usepackage{enumitem}
\usepackage{bm}
\begin{document}  

\title{Can we distinguish low mass black holes in neutron star binaries?
     } 
{
    \author{Huan Yang${}^{1,2}$, William E. East${}^1$, and Luis Lehner${}^1$}
\affil{
${}^1$Perimeter Institute for Theoretical Physics, Waterloo, Ontario N2L 2Y5, Canada\\
${}^2$University of Guelph, Guelph, Ontario N2L 3G1, Canada
}
\begin{abstract} 
The detection of gravitational waves from coalescing binary neutron stars
represents another milestone in gravitational-wave astronomy. However,
since LIGO is currently not as sensitive to the merger/ringdown part of
the waveform, the possibility that such signals are produced by a black
hole-neutron star binary can not be easily ruled out without
appealing to assumptions about the underlying compact object populations. We review a few astrophysical
channels that might produce black holes below 3 $M_{\odot}$ (roughly the
upper bound on the maximum mass of a neutron star), as well as existing
constraints for these channels. We show that, due to the uncertainty in the
neutron star equation of state, it is difficult to distinguish
gravitational waves from a binary neutron star system, from those of a
black hole-neutron star system with the same component masses, assuming
Advanced LIGO sensitivity. This degeneracy can be broken by accumulating
statistics from many events to better constrain the equation of state, or
by third-generation detectors with higher sensitivity to the late spiral to
post-merger signal. We also discuss the possible differences in
electromagnetic counterparts between binary neutron star and low mass black
hole-neutron star mergers, arguing that it will be challenging to
definitively distinguish the two without better understanding of the underlying
astrophysical processes.
\end{abstract}

\maketitle 

\section{Introduction}

Merging binary neutron stars have just resoundingly
been shown to produce both strong gravitational wave (GW) signals, and copious
electromagnetic (EM) emission covering a large frequency range by the
recent event GW170817~\citep{PhysRevLett.119.161101,2041-8205-848-2-L12,2041-8205-848-2-L13,coulter2017swope,troja2017x,2041-8205-848-2-L21} (see also, e.g.
\cite{kochanek1993gravitational,li1998transient,rosswog2005mergers,metzger2010electromagnetic,Murguia-Berthier:2014pta,Metzger:2011bv,ando2013colloquium}).
The joint
observation of GWs with, for instance, gamma-ray bursts, x-rays, ultraviolet/optical/infrared
transients, or radio afterglows is now beginning to provide unprecedented information about
the violent dynamics of hot, dense nuclear matter under extreme gravity.  With
three gravitational wave detectors now online, the ability to localize the
source of GW events (albeit within a still rather large window) facilitates
identifying EM counterparts~\citep{PhysRevLett.119.161101,Abbott:2017oio}.

It is natural to assume a compact object is a neutron star (NS), instead of a
black hole (BH), if its mass is below the upper bound of a non-rotating NS.
Such an assumption has been also supported by the observed mass distribution of
NSs and BHs in binaries, and applied to distinguish NS-NS and BH-NS binaries
through the use of component mass measurement to identify a possible ``mass gap"
in BHs \citep{hannam2013can,mandel2015distinguishing,littenberg2015neutron}.
However, the advent of GW astronomy allows us the chance to re-examine preconceptions
that might be biased by previously available observations~\footnote{Arguably this has
already taken place with GW observations revealing the existence of BHs with
masses $>20M_{\odot}$ \citep{TheLIGOScientific:2016htt}.}.  In this paper, we consider the possible
existence of low-mass black holes (LMBHs) with a mass range that overlaps that
of normal NSs. We briefly review possible formation channels of such BHs, determine the
prospects for identifying them through GW and/or multi-messenger detections,
and discuss the implications upon detecting such objects. 

\section{LMBH Formation channels}

Here we list several possible formation
channels to generate LMBHs with masses $<3\ M_{\odot}$.  First, stellar-mass
BHs could come from primordial density fluctuations. In the range we are
considering ($\sim 1-3\ M_\odot$), existing constraints stem from microlensing
measurements \citep{carr2016primordial} indicating that their mass fraction
compared to dark matter is $f \le 5\%$.   We
expect that the ratio between such LMBHs and normal NSs in a galaxy to be
\begin{align}
f \frac{M_{\rm total} \Omega_{\rm DM}}{N_{\rm NS} M_{\rm NS}} \sim 330 \frac{M_{\rm total}}{10^{12} M_\odot} \frac{f}{5\%} \left (\frac{N_{\rm NS}}{10^8} \frac{M_{\rm NS}}{1.3 M_\odot} \right )^{-1} \frac{\Omega_{\rm DM}}{0.845}\,,
\end{align}
for a Milky Way-like galaxy (the Milky way values for the total mass  $M_{\rm total}$ and number of NSs $N_{\rm NS}$ are estimated in \cite{dehnen1998mass} and \cite{camenzind2007compact}), where $\Omega_{\rm DM}$ is the mass faction of dark matter in total matter density. With this upper bound saturated, if the cross section for dynamically
capturing a NS is approximately the same as the one for a BH
with similar mass---which should be the case since such capture is dominated
by GW emission \citep{East:2011xa}, it is possible that the merger rate of LMBH-NS binary is
actually greater than the merger rate of dynamically formed NS-NS binaries.  Similarly, motivated by the discussion in
\cite{capela2013constraints} and \cite{fuller2017primordial}, NSs, white dwarfs,
or even main sequence stars could capture mini-primordial black holes (PBHs)
\footnote{Here we are referring to PBHs with masses much smaller than one solar mass.}
causing most of the star's material to be accreted to produce a final BH 
with stellar mass. It is, however, not clear what fraction of NSs could become 
LMBHs through this process. In fact, a bound on the mini-PBH population was
obtained in \cite{capela2013constraints} assuming that not all NSs are destroyed by
PBH captures.  It has also been proposed that asymmetric dark matter could
accumulate in centers of NSs through nucleon scattering, and
eventually form a seed BH~\citep{Goldman:1989nd,Bramante:2014zca,Bramante:2017ulk}, providing another
scenario for converting a NS to a BH of similar mass.

Another possible way to produce LMBHs is through 
a supernovae explosion, a standard
mechanism for creating compact objects.
If the explosion is driven by 
 rapidly growing instabilities (10-20 ms) black hole with masses $> 5\ M_{\odot}$ are
expected, but slow ones can produce lower masses~\citep{2012ApJ...757...91B}. To date, observations
point to the former option, but the existence of a ``mass gap'' is by no means a definitive fact \citep{kreidberg2012mass}.

Additionally, it is also possible that the final BH produced by NS-NS merger could become
subsequently captured in a new LMBH-NS binary. Such hierarchical mergers were
discussed in \cite{fishbach2017ligo} and \cite{gerosa2017merging} from the detection
perspective, and in \cite{antonini2016merging} as a way to estimate the rate of
of BH-BH mergers.
Similarly, NSs could gain mass through accretion and collapse to a BH falling in the mass range considered here \citep{nakamura1983general,Vietri:1999kj,MacFadyen:2005xm,Dermer:2006pw}.
Because of the uncertainty in upper threshold mass of a normal NS, LMBHs formed through NS accretion-induced collapse or collisions can not be easily 
distinguished from candidates in other channels through mass measurement.

We also note that the range of BH masses allowed by the above channels could
also be modified by possible departures from General Relativity, the existence of
additional fields in nature, and/or exotic compact objects (e.g. \cite{kaup1968klein,cardoso2016gravitational, mendes2017tidal,Liebling:2012fv}) whose dynamics can
yield LMBHs through collapse or mergers. Consequently, through dynamical captures, LMBH-NS
systems could be produced with the masses achievable in each possible scenario.
Naturally, with such range of possible formation channels together with current 
uncertainties as to their likelihood, the rate of LMBH-NS binaries is unknown. \footnote{{There are studies on merger rates of different types of compact binaries (BH-BH, NS-NS, NS-BH) based on various population synthesis models (e.g. \cite{belczynski2016compact}).}} Thus, future gravitational
wave observations of such systems (possibly requiring electromagnetic counterparts) will be key to understanding
this theoretically possible population.


\section{Degeneracy of tidal effects}

We argue that the leading order tidal effects on the GW signal of an
inspiraling compact object binary are in fact degenerate between a
NS-NS and a BH-NS binary, when considering different equations of state
(EOSs) of the star (and hence setting its radius, etc.).
We begin by noting that the phase of the inspiral waveform as a function of
frequency can be written as~\citep{vines2011post,Sennett:2017etc}
\begin{align}
\Psi(f) = & \frac{3}{128 (\pi \mathcal{M} f)^{5/2}} [1+\alpha_{\rm 1PN} x+... \nonumber \\
&+(\alpha_{\rm 5PN}+\alpha_{\rm tide})x^5+...]
\end{align}
where $x=(\pi M f)^{2/3}$, $\mathcal{M}=(m_1m_2)^{3/5}/M^{1/5}$, and $M=m_1+m_2$ is the total mass.
The $\alpha_{\rm 1PN}$, \ldots, $\alpha_{\rm 5PN}$ terms encode the various order Post-Newtonian (PN)
effects, while the leading order tidal correction is given by
\begin{align}
    \alpha_{\rm tide} = -24\left [ \left ( 1+12 \frac{m_2}{m_1}\right ) \frac{m^5_1}{M^5}\Lambda_1+(1 \leftrightarrow 2)\right ]\,,
\end{align}
where $\Lambda_1$ and $m_1$ are the dimensionless tidal deformability parameter
(normalized by mass to the fifth power) and mass of the first compact object (which here, we will
always assume is a NS), respectively, and the second term exchanges these quantities for that
of the second compact object (which we take to either be a NS or a BH).  It
follows from the above that as long as the NS in a BH-NS binary satisfies an
EOS that has tidal deformability
\begin{align}\label{eqm}
\Lambda'_1 = \Lambda_1 +\frac{m_2+12 m_1}{m_1+12m_2} \frac{m^5_2}{m^5_1} \Lambda_2\,,
\end{align}
we can not distinguish its inspiral waveform from that of a NS-NS
system with $(\Lambda_1,\Lambda_2)$ for the respective stars, up to the leading
PN order in tidal corrections.  

We can illustrate that this leading order difference in the tidal effects of a
NS-NS versus BH-NS system can be readily accommodated into uncertainties in the
EOS. To give a concrete example of this, we consider a one parameter family of
EOSs given by the SLy equation~\citep{Douchin:2001sv} of state at low densities,
and a $\Gamma=3$ polytrope at high densities, also roughly consistent with SLy,
where the parameter sets the pressure at some reference
density~\citep{Read:2008iy} (we consider $P=10^{34.1}$ to $10^{35.1}$
dyne/cm$^2$ at $\rho=5\times10^{14}$ gm/cm$^3$). Then, for a given set of binary
parameters, we find the mapping between equations of state in this family such
that Eq.~\ref{eqm} is satisfied (see~\cite{Gagnon-Bischoff:2017tnz} for details
on computing Love numbers). We show this mapping, in terms of the
amount by which the NS radius in the BH-NS binary has to be larger, relative to
the radius of the corresponding NS in a NS-NS binary, in Fig.~\ref{fig:radius}.
Typically this increase is less than 2 km. Furthermore, since the tidal effects
in a binary neutron star are dominated by the star of larger radius, the required increase
can be quite small when the smaller object is taken to be the black hole.  We
note in passing that attributing measured tidal effects to a BH-NS binary
generically implies a stiffer EOS, and so could be favored, if the softer EOS
implied by a NS-NS binary is in tension with other observations (e.g., of the
maximum allowed NS mass).

In addition, this mapping assumes we know the component masses exactly, and
the uncertainty just lies in the EOS. If we also fold in the uncertainty in
component masses \citep{hannam2013can,chatziioannou2014spin}, there is a greater degeneracy.

\begin{figure}
\begin{center}
\includegraphics[width=\columnwidth,draft=false]{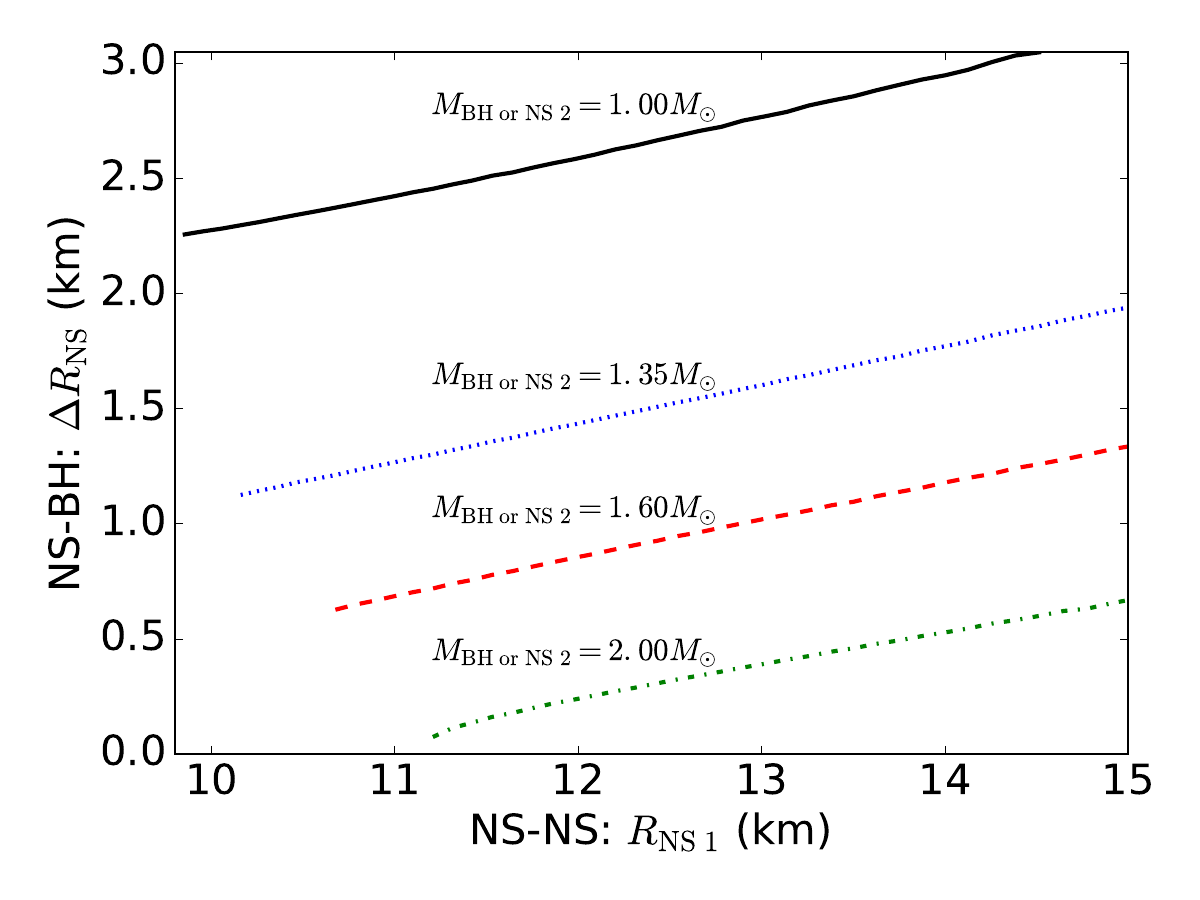}
\end{center}
\caption{ The amount by which the NS radius has to be increased---relative to
    the radius of the corresponding NS in a NS-NS binary with the same masses---,
    for a BH-NS binary to have the same leading order tidal effects.  
     In all
    cases, one NS (labelled NS 1) is assumed have a mass of $1.35\ M_{\odot}$, while the
    different curves correspond to a BH or second NS with a mass ranging from 1
    to 2 $M_{\odot}$.  
    The leftmost value of each curve (smallest value of $R_{\rm NS\ 1}$) corresponds to the point where for the family of EOSs considered here,
    the BH-NS EOS is no longer consistent with a maximum NS mass greater
    than $2\ M_{\odot}$. Since the corresponding NS-NS EOS is softer, the constraint that the NS-NS EOS
    be consistent with a maximum NS mass greater than $2\ M_{\odot}$ is stricter, and requires $R_{\rm NS\ 1}\gtrsim 11.2$ km.
\label{fig:radius}
}
\end{figure}

\section{Prospects for GW detection}

As discussed above, the leading
order tidal effect is degenerate between LMBH-NS and NS-NS systems, as long as
there is sufficient uncertainty in EOS to allow the mapping in Eq.~\eqref{eqm}. This
degeneracy may be resolved in several ways: through the measurement of
the next-to-leading PN order tidal effects, as they contain different mass and frequency 
dependence; through the difference between two types of waveforms in the
late-inspiral stage, where the tidal disruption of the NS strongly influences the
waveform; or through the accumulation many NS-NS events and the consequential reduction
in the uncertainty of the star's EOS (notably its radius), to break the degeneracy.
Since the effect of PN corrections to tidal effects is $\sim10\%$--$20\%$~\footnote{It was estimated in \cite{hinderer2010tidal} that the PN correction
contributes approximately $10\%$ modification to the tidal effect at $450 {\rm
Hz}$. Roughly speaking, to detect an effect that is ten times smaller in
amplitude with similar accuracy or confidence, we need ten times enhancement in
SNR. See also~\cite{vines2011post}.}, we focus on the latter two possibilities here. 

In order to
distinguish two waveforms $h_{\rm NSNS}$ and $h_{\rm BHNS}$, we adopt the measure
\begin{align}\label{eqsnrd}
{\rm SNR}^2_{\rm \Delta} = 4\int df \frac{|h_{\rm NSNS}(f)-h_{\rm BHNS}(f)|^2}{S_n(f)}\,,
\end{align}
with $S_n(f)$ being the spectra density of Advanced LIGO detector noise. If  ${\rm SNR}_{\rm \Delta} \ge 1$, we shall say that the two waveforms are marginally distinguishable \citep{lindblom2008model}. This threshold has 
to be raised if we require higher statistical significance. 
The inspiral signals of LMBH-NS and NS-NS waveforms terminate at different
respective characteristic frequencies. For a LMBH-NS system, the signal
terminates at the cut-off frequency $f_{\rm cut}$, which is related to the
tidal disruption of NS within a BH-NS binary.  For simplicity, in the following estimate 
we will assume that in LMBH-NS binaries when the star is disrupted, the gravitational wave is negligible,
and ignore the post-merger part of the waveform for both types of systems. 
Since Advanced LIGO/VIRGO's sensitivity degrades considerably
at the high frequencies where contact (for NS-NS systems) or disruption (for LMBH-NS systems) occurs, a rather
good approximation to ${\rm SNR}^2_{\rm \Delta}$ can be readily obtained this way.
Based on
\cite{shibata2009gravitational}, $f_{\rm cut}$ is actually much higher than the
frequency that the NS undergoes mass shedding, and it depends on the mass ratio
of the system and the NS EOS (in particular the NS compactness $\mathcal{C} \equiv M_{\rm NS}/R_{\rm NS}$).  Here,  
we adopt the fitting formula in \cite{pannarale2015gravitational} \footnote{We
note that this fitting formula was obtained using a set of data where the
minimum mass ratio was $2$. This means that Fig.~\ref{fig:snr} may contain
systematic error due to extrapolation. In fact, there are limited simulations
available for the low mass-ratio systems considered here, which motivates 
future numerical studies within this parameter range.} (under a simplified
assumption that the BH is nonspinning)
\begin{align}
f_{\rm cut} = \frac{1}{M} \sum_{i,j} f_{ij} \mathcal{C}^i Q^j\,,
\end{align}
where $Q=M_{\rm BH}/M_{\rm NS}$ is the mass ratio and $f_{ij}$ are numerical
coefficients given in \cite{pannarale2015gravitational}.  

For a NS-NS system, the
inspiral ends at the contact frequency $f_{\rm contact}$ of the two NSs \citep{PhysRevD.85.123007}:
\begin{align}
f_{\rm contact} =\frac{1}{\pi M} \left ( \frac{m_1}{M} \frac{1}{\mathcal{C}_1}+\frac{m_2}{M} \frac{1}{\mathcal{C}_2}\right )^{-3/2}\,.
\end{align}

Eq.~\eqref{eqsnrd} can then be approximated by, 
\begin{align}\label{eq:snrp}
{\rm SNR}^2_{\rm \Delta} \approx 4\int^{f_{\rm max}}_{f_{\rm min}} \, df \frac{|h_{\rm 3PN}(f)|^2}{S_n(f)}\,,
\end{align}
where we have used a sky-averaged $3{\rm PN}$ inspiral waveform
\citep{kidder2008using} in this range. In Fig.~\ref{fig:snr}, 
we plot ${\rm SNR}_{\rm \Delta}$ as a function of the mass ratio of the binary,
and the radius of the lighter NS. 
We assume that the binary is at a distance of $50\ {\rm Mpc}$,
and the mass of the lighter NS is assumed to be $1.35\ M_\odot$.
Within most of the parameter regime we consider here, ${\rm SNR}_{\rm \Delta}$ is
bounded below $\sim 1$. The regime with very low ${\rm SNR}_{\rm \Delta}$
corresponds to the cases with $f_{\rm cut} \approx f_{\rm merger}$.
Nevertheless, the distinguishability will be improved if the source is
closer, or if we include the post-merger part of the waveform (regarding which there  
are still significant modelling uncertainties). For example, the SNR of the full
post-merger waveform is estimated to be around $1.5$ \citep{Clark:2015zxa} for a
$1.35\ M_\odot+1.35\ M_\odot$ NS binary at a distance $d=50\ {\rm Mpc}$ and with
the ``TM1" EOS, while the SNR contribution from the dominant mode is significantly
lower \citep{yang2017gravitational}, depending on the EOS.

\begin{figure}
\begin{center}
\includegraphics[width=\columnwidth,draft=false]{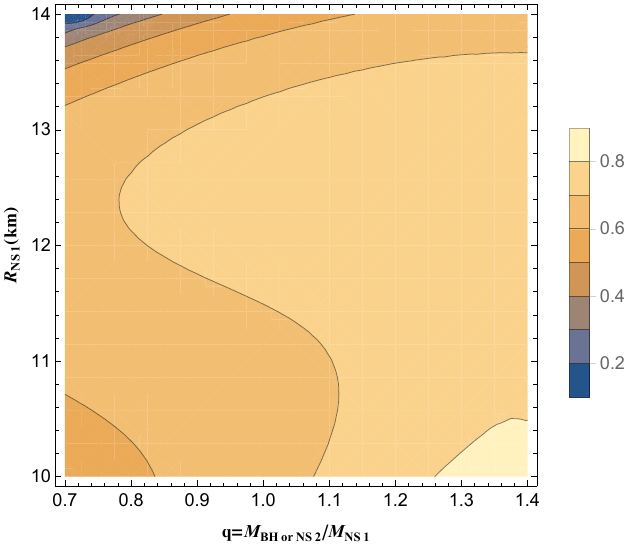}
\end{center}
\caption{ The SNR computed using Eq.~\eqref{eq:snrp} and assuming a binary at a distance of $50\ {\rm Mpc}$. The mass of NS $1$ is assumed to be $1.35 M_\odot$. This SNR is a function of the radius of the NS $1$ and the binary mass ratio. { As the SNR is inversely proportional to distance, a $1.35 M_\odot+1.35 M_\odot$ (with NS radius $\sim 12 {\rm km}$) binary event in the Virgo cluster should have ${\rm SNR}_\Delta$ around $2$.}}
    
\label{fig:snr}

\end{figure}

On the other hand, multiple detections of NS-NS mergers will be able to
constrain the NS EOS, which could help break the degeneracy indicated by
Eq.~\eqref{eqm}. According to \cite{hinderer2010tidal}, a single Advanced LIGO
event from a distance $d$ typically  constraints $\Lambda$ to an accuracy
\begin{align}
\Delta \Lambda \sim & 2.9\times 10^{3}  \nonumber \\
&\left ( \frac{M}{2.7\ M_\odot}\right )^{-2.5} \left ( \frac{m_2}{m_1}\right )^{0.1} \left ( \frac{f_{\rm end}}{{\rm 500\ Hz}}\right )^{-2.2} \left ( \frac{d}{100\ {\rm Mpc}}\right )\,.
\label{eqn:delta_lam}
\end{align} 
Assuming a similar end frequency for integration $f_{\rm end} \sim 500\ {\rm
Hz}$ as in~\cite{hinderer2010tidal} \footnote{The quantity $f_{\rm end}$ is set
to avoid higher PN tidal effects and nonlinear hydrodynamical coupling. In
\cite{hinderer2010tidal}, the fitting formula was obtained by choosing $400
{\rm Hz} \ge f_{\rm end} \ge 500 {\rm Hz}$. The end frequency adopted in ~\cite{PhysRevLett.119.161101} is the frequency of Innermost-Stable-Circular-Orbit, which is above 1 kHz.}, and an equal mass binary with $M
\sim 2.7\ M_\odot$, we obtain an estimate on the tidal deformability of a star:
$\Delta \Lambda_1 (1.35 M_\odot) \sim 2.9\times 10^3$. 
Assuming a low-spin prior, GW170817 places on upper bound $\Lambda_1(1.4 M_{\odot})\leq 800$
at the $90\%$ confidence level, and an upper bound of $\leq 1400$ if the prior is relaxed
to allow for high spin~\citep{PhysRevLett.119.161101}. Such constraints on $\Lambda_1$ also limit the allowed radius 
of neutron star given the family of EOS assumed in this paper.
In Fig.~\ref{fig:delta_lambda}, we compare this uncertainty in the tidal deformability
to the amount by the tidal deformability has to been changed in order for BH-NS
binary to have the same leading order tidal effects as a corresponding NS-NS binary
with the same masses.

With $N$ identical
detections, and under the same high-spin prior, such uncertainty scales as $\Delta_N \Lambda_1 (1.4 M_\odot) \sim
1.4 \times 10^3 N^{-1/2}$. In reality, the component masses and source
distances are different for different events, and it is possible that the best
event of $N$ previous detections dominates the constraint of NS EOS. 

\begin{figure}
\begin{center}
\includegraphics[width=\columnwidth,draft=false]{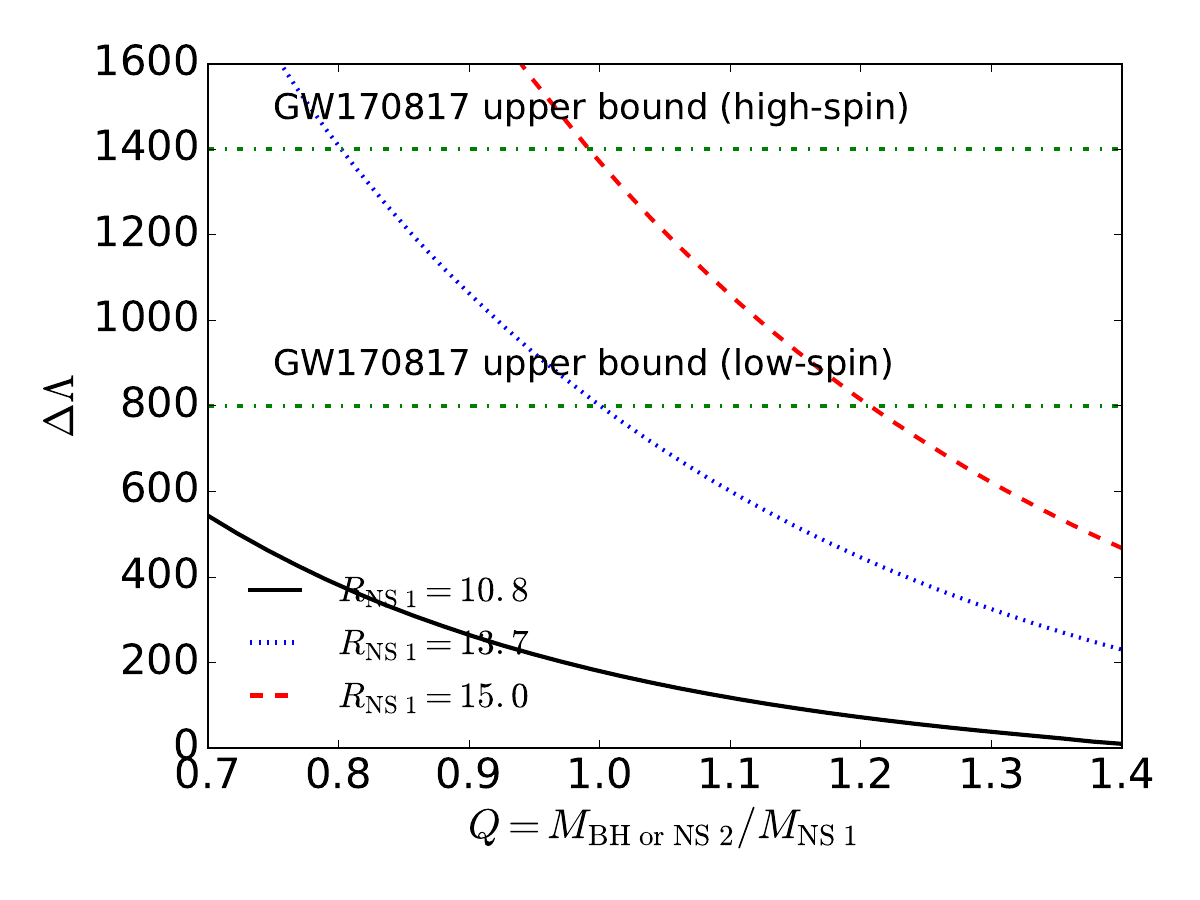}
\end{center}
\caption{ The amount by which the dimensionless tidal deformability $\Lambda$
    has to be increased, relative to the quantity of the corresponding NS in a
NS-NS binary with the same masses, for a BH-NS binary to have the same leading
order tidal effects.  This is shown as a function of mass ratio.  In all cases,
one NS is assumed have a mass of $M_{\rm NS\ 1}=1.4\ M_{\odot}$.  The
different curves correspond to different choices of EOS for the NS-NS, and
hence the different values of the radius of $1.4\ M_{\odot}$ NS in the NS-NS,
as shown in the legend.  For comparison, we also show the upper-bound
constraints on $\Lambda$ from GW170817 assuming either a high-spin prior, or a
low-spin prior on the source. The highest NS radius shown (dotted red curve)
saturates the former constraint on $\Lambda$, while the middle value for
NS radius (dotted blue curve) saturates the latter constraint. 
    }
\label{fig:delta_lambda}
\end{figure}

In the above discussion, we have not accounted for the effect of spin, which is
particularly important for LMBHs formed through hierarchical mergers, as they
are expected to have relatively high spin ($a \sim 0.7$)
\citep{fishbach2017ligo,gerosa2017merging} \footnote{The effective spin of a
most rapidly-spinning pulsar known so far is $\sim 0.4$
\citep{hessels2006radio}.}. Such a spin magnitude will generate $\sim
0.15$ mismatch between a non-spinning BH-NS waveform and a generic precessing 
waveform if the mass ratio is around $2$ (see Fig.~8 of
\cite{harry2014investigating}). The relation between distinguishable mismatch
and SNR is discussed in \cite{baird2013degeneracy}.


\section{Multi-messenger detection}

An important question is whether
multi-messenger signals can help us to identify a LMBH. In other
words, what are the possible features of LMBH-NS systems that distinguish them
from NS-NS systems, besides direct GW observation of the merger waveform?
(BH-BH systems in stellar mass ranges are not expected to produce EM signals,
so the clear presence of such signal would favor a system with at
least one neutron star.) We argue that current limitations in our theoretical 
understanding of the underlying astrophysical process giving rise to electromagnetic
counterparts make it difficult to clearly distinguish a binary with only one neutron star 
versus  a binary with two neutron stars. In what follows, we discuss
several leading counterpart prospects (see also e.g.~\cite{Metzger:2011bv}) but note that the era of multimessenger astronomy
will bring an increased understanding of them, as well as awareness of further ones.

Several EM counterparts have been proposed that occur within (tens of) milliseconds prior to merger,
including possible emissions related to crust-cracking due to tidal
effects~\citep{2012PhRvL.108a1102T} (with associated luminosities which
could reach levels of order $L\approx 10^{48}$ erg/s)
and magnetosphere interactions~\citep{McWilliams:2011zi,Palenzuela:2013hu,Metzger:2016mqu,Piro:2012rq}
(with associated luminosities which could reach levels of order $L\approx 10^{43} (B/10^{14}\ G)^2$ erg/s). 
However, uncertainties in the EOS and magnetization level of the NS makes
distinguishing such signals seem unlikely.

As the merger proceeds, the star will be disrupted by the LMBH and give rise
promptly to an accreting black hole---the most popular central engine model for
a short gamma-ray burst (sGRB). On the other hand, binary neutron stars can
themselves power a jet which, as discussed in~\cite{Murguia-Berthier:2014pta},
can escape if the jet breaks in a sufficiently short time. Thus, a sGRB seen to
take place nearly coincident with the peak in GWs would not provide a clear
discerning prospect. It is important to keep in mind that the newly formed
massive neutron star will reach very high
magnetizations ($B \approx 10^{15-17}\ G$ \citep{Anderson:2008zp,Kiuchi:2017zzg}. If it collapses, large
amounts of energy ($L\approx 10^{49} (B/10^{15}\ G)^2$ erg/s) 
could be released rather isotropically~\citep{Lehner:2011aa}
setting the stage for possible less intense high energy (gamma, x-ray) emissions which do not
necessarily require the observer to be specially aligned. Interestingly, the (short) GRB170817A
associated with the gravitational wave event GW170817 is less luminous than typical sGRBs~\citep{2041-8205-848-2-L13}. This fact
could be explained by the viewing angle but also through different burst mechanisms/models. 

The collapse of a hypermassive NS to a BH, however, can take place in a
significantly delayed fashion, and the resulting accreting BH state would fit naturally
in the ``canonical picture'' of an accreting BH launching the jet, 
especially if the jet is Poynting flux dominated.
It is tempting to speculate that, in such a paradigm, a significantly delayed sGRB would 
favor a NS-NS system (as timescales for launching a jet could be around $\approx 100$ ms,
e.g.~\cite{Paschalidis:2014qra,Ruiz:2016rai}); however, the time required to set the right
topology and strength of the magnetic fields required for 
launching a jet~\footnote{Recall that tidal disruption induces a
mainly toroidal field configuration at first~\citep{2010PhRvL.105k1101C}.}
(e.g.~\cite{2009MNRAS.394L.126M}) introduces a delay that can potentially blur
the differences between a BH-accretion scenario set up promptly after the
merger (through BH-NS or NS-NS mergers) or delayed (via a NS-NS merger that
produces a long-lived remnant). Furthermore, current uncertainties in key effects like
effective viscosity of the forming disk, magnetization levels of the star, accretion characteristics,
as well as the sGRB model itself (e.g.~\cite{Narayan:2001qi,Piran:2004ba}) currently stand 
in the way of clearly distinguishing
the progenitors based on such a delay.

Another way in which a LMBH-NS may potentially differ from a NS-NS binary is in the
amount (and neutron richness) of NS material that is unbound during the merger. This ejecta will
undergo r-process nucleosynthesis, building up heavy elements that decay,
powering a so called kilonova/macronova~\citep{Li:1998bw,2005astro.ph.10256K,Metzger:2016pju}.
In a rather spectacular fashion, such observations have been identified as counterparts 
to GW170817, e.g.~\cite{Smartt:2017fuw}, \cite{Drout:2017ijr}.
The greater the mass $M_{\rm ej}$ of material that is ejected, the brighter the
transient, and the longer the timescale on which it will peak.  In terms of the
velocity of the ejecta, such an EM transient will peak on
timescales $t_{\rm peak}\sim0.3 (M_{\rm ej}/0.01\ M_{\odot})(v/0.2c)$ days, in
the ultraviolet/optical to near infrared frequencies, with peak luminosities of
$L\sim1.6\times10^{41}(M_{\rm ej}/0.01\ M_{\odot})(v/0.2c)$
ergs/s~\citep{2013ApJ...775...18B}.  On longer timescales of 
$\sim 2.6 (E_{\rm ej}/10^{50}{\rm \ erg})^{1/3}(v/0.2c)^{-5/3}$ years (where $E_{\rm ej}$ 
is the kinetic energy of the ejecta), there may
also be a radio transient associated with the collision of this material with
the interstellar medium~\citep{2011Natur.478...82N,Hallinan:2017woc}.  

Simulations of NS-NS mergers typically find ejecta of $\lesssim 0.01\ M_{\odot}$,
with the most ejecta coming from mergers with soft EOSs. With
unequal mass ratios~\citep{Lehner:2016lxy,Sekiguchi:2016bjd}, the ejected material is
highly neutron rich, and the amount is on the higher end across EOSs.  Higher mass-ratio 
simulations of BH-NS mergers find significant ejecta when the black hole has
non-negligible spin aligned with the orbital angular momentum and/or the NS has a
larger radius~\citep{Foucart:2014nda,Kawaguchi:2016ana}, in which case the
amount of ejecta can be up to $\sim0.1M_{\odot}$. 
Hence, an unusually bright ejecta-powered
transient would seem to favor a LMBH-NS merger, though a transient consistent
with $\lesssim0.01\ M_{\odot}$ ejecta could be attributed to either.  
We note, however, 
BH-NS mergers with nearly
equal masses are not well studied (see~\cite{Etienne:2007jg,Shibata:2009cn} for
some early studies), and further scrutiny will be required to delineate their
properties across parameter space.  

An additional caveat to the above discussion is that non-negligible NS spin, on the order of $a\sim0.1$, has also
been shown to enhance the amount of ejecta to the level of a few percent of a
solar mass~\citep{East:2015vix,East:2016zvv,Dietrich:2016lyp} (this falls within the allowed
range for the spin along the orbital angular momentum estimated in GW170817, notice the component
orthogonal to it is however not constrained).  Orbital
eccentricity at merger can also significantly increase the
amount of ejecta~\citep{East:2012ww,Radice:2016dwd}, though presumably this will be well
constrained by the GW signal.

\section{Conclusion}

 It is conceivable that LMBHs may be produced
through PBH capture, supernovae, NS-NS mergers, the collapse of exotic compact
objects, or other such phenomena. Therefore, their existence is tightly
connected to the astrophysical population/distribution of these seeding objects and
the underlying fundamental physics that governs them. Because of the
uncertainty in the NS EOS and the degeneracy in tidal effect of LMBH-NS and NS-NS
systems in the inspiral stage, it appears challenging for Advanced LIGO to
definitively identify such objects. The ability to differentiate between the two can be improved by
better understanding their respective post-merger waveforms, as well as
achieving better GW detector sensitivity \citep{miao2014quantum,miao2017towards} and
accumulating statistics from many detections
\citep{yang2017gravitational,bose2017neutron,yang2017black}.  The similarities
in the potential EM counterparts to the two systems, within theoretical
uncertainties, also makes distinguishing them with multimessenger astronomy
challenging, and calls for a better understanding of the underlying
astrophysical processes.  Such a task of refining models and honing in on the
relevant parameter space will benefit tremendously from a dialogue with
observations as they take place. 

If such a LMBH were discovered, the problem of identifying its formation channel
would naturally arise. One possible indicator could be the spin of the
LMBH---one can compute its prior distributions in each formation channel and
compare them with the posterior distributions of each detection.  The mass and
redshift information of these objects may also help distinguish their origins.
Excitingly, third-generation GW detectors will be capable of detecting non-vacuum compact
binary mergers up to $z \sim 6$ \citep{abbott2017exploring}. If LMBHs are present even
in a small portion of such mergers, they will guide fruitful discoveries in physics and astronomy.

\acknowledgments
We thank J\'{e}r\'{e}mie Gagnon-Bischoff and Nestor Ortiz
for sharing their code for computing tidal Love numbers. This research was supported in
part by NSERC, CIFAR, and  by the Perimeter Institute for Theoretical
Physics.  Research at Perimeter Institute is supported by the
Government of Canada through the Department of Innovation, Science and
Economic Development Canada, and by the Province of Ontario through
the Ministry of Research and Innovation.

\bibliographystyle{apj}
\bibliography{master}
\end{document}